\documentclass[prd,email,twocolumn,showpacs,showkeys,preprintnumbers,amsmath,amssymb,nofootinbib]{revtex4-1}

\def\[{\left\lbrack}
\def\]{\right\rbrack}

\def\({\left(}
\def\){\right)}
\newcommand{\be}{\begin{equation}}
\newcommand{\ee}{\end{equation}}
\newcommand{\ea}{\end{eqnarray}}
\newcommand{\ba}{\begin{eqnarray}}

\usepackage{color}
\begin{document}

\title{Obtaining non-Abelian field theories via Faddeev-Jackiw symplectic formalism}


\author{E. M. C. Abreu$^a$} 
\email{evertonabreu@ufrrj.br}
\author{A. C. R. Mendes$^b$}
\email{albert@fisica.ufjf.br}
\author{C. Neves$^c$}
\email{clifford.neves@gmail.com} 
\author{W. Oliveira$^b$} 
\email{wilson@fisica.ufjf.br}
\author{R. C. N. Silva$^b$}
\email{rodrigocnsilva@yahoo.com.br}
\author{C. Wotzasek$^{d}$}
\email{clovis@if.ufrj.br}

\affiliation{${}^{a}$Grupo de F\' isica Te\'orica e Matem\'atica F\' isica,
Departamento de F\'{\i}sica, Universidade Federal Rural do Rio de Janeiro\\
BR 465-07, 23890-970, Serop\'edica, Rio de Janeiro, Brazil\\
${}^{b}$Departamento de F\'{\i}sica,ICE, Universidade Federal de Juiz de Fora,\\
36036-330, Juiz de Fora, MG, Brazil\\
${}^{c}$Departamento de Matem\'atica e Computa\c{c}\~ao, Universidade do Estado do Rio de Janeiro\\
Rodovia Presidente Dutra, km 298, 27537-000, Resende, Rio de Janeiro, Brazil\\
${}^{d}$Instituto de F\'{\i}sica, Universidade Federal do Rio de Janeiro\\
Caixa Postal 68528, 21945-970, Rio de Janeiro, Brazil\\
\bigskip
\today}

\pacs{11.15.-q; 11.10.Ef; 02.20.Sv}

\keywords{non-Abelian field theories, Faddeev-Jackiw symplectic formalism}

\begin{abstract}
\noindent In this work we have shown that it is possible to construct non-Abelian field theories employing, in a systematic way, the Faddeev-Jackiw symplectic formalism. This approach follows two steps. In the first step, the original Abelian fields are modified in order to introduce the non-Abelian algebra.  
After that, the Faddeev-Jackiw method is implemented and the gauge symmetry relative to some non-Abelian symmetry group, 
is introduced through the zero-mode of the symplectic matrix. We construct the $SU(2)$ and $SU(2)\otimes U(1)$ Yang-Mills theories having as starting point the $U(1)$ Maxwell electromagnetic theory. 
\end{abstract}

\maketitle

\pagestyle{myheadings}
\markright{Obtaining non-Abelian field theories via Faddeev-Jackiw symplectic formalism}

\section{Introduction}

The consistent quantization method for constrained systems was introduced by Dirac \cite{Dirac}. In particular, the Dirac formalism analyzes the canonical structures of theories, which are essential to the development of quantum theories. 

Some years ago Faddeev and Jackiw (FJ) \cite{Jackiw} suggested a symplectic approach to constrained systems based on a first-order constructed Lagrangian.   Described firstly as a ``Hamiltonian formulation" by the authors, it is an alternative kind of canonical quantization method to the standard Dirac procedure, since this last one classifies the constraints in first or second class, primary or secondary.  The FJ formalism avoids the definition of unnecessary variables like the momenta and introduces the so-called $2n$-component phase-space coordinates, which encompass all the coordinates and the momenta of the original theory.  Finally, as we will see, in the FJ technique, a one-form Lagrangian is constructed together with a symplectic $2n\times2n$ matrix.  Another feature of the FJ method is that the physical degrees of freedom are pinpointed without fixing the gauge.  Later on, Floreanini and Jackiw promoted the canonical quantization \cite{floreanini}.  The path integral approach was developed recently \cite{lh}.

After the work of Faddeev and Jackiw, Barcelos-Neto and Wotzasek (BW) \cite{Wotzasek} extended the FJ symplectic formalism for the case where the constraints are not completely eliminated.  In their work the constraints produce deformations in the symplectic two form matrix in such a way that, when all constraints are considered, the symplectic matrix becomes invertible. As a result, they have directly obtained the Dirac brackets from this inverse. It is important to mention that, when the two form matrix is singular then no new constraint is obtained from the corresponding zero-mode. This is the case when one deals with gauge theories. At this point one introduces the gauge condition we want as a constraint and the two form matrix becomes invertible.

In recent works \cite{Oliveira}, the FJ symplectic formalism has been used in a systematic way for different purposes: the study of hidden symmetries, the construction of equivalent gauge theories (duality), noncommutative field theory,  to solve the obstruction problem to the construction of the canonical Lagrangian formulation for rotational systems and other.

The purpose of the present work is to show a completely new application of the FJ symplectic formalism.  We will demonstrate exactly that the method can also be used to obtain, in a systematic way, non-Abelian field theories. The systematic method that we will present here to construct non-Abelian field theories has the following methodology. Firstly, the original Abelian fields are changed in order to introduce the non-Abelian algebra.  After that, the FJ symplectic method is implemented and the gauge symmetry, that belongs to some  non-Abelian symmetry group, $SU(2)$ for instance, is introduced. This is accomplished using the zero-mode of the symplectic matrix, which must be considered singular. Thus, it is possible to calculate the new one-form tensor and the canonical momenta.  Notice that new terms will arise due to the new symmetry group considered.

Again, it is very important to notice that, although the results are obviously not new, what is new is the possibility that the FJ method can be used to construct such mapping which begins with an Abelian theory and discloses non-Abelian features and finishes with the well known non-Abelian theory.  We believe that this new approach involving the FJ formalism increase remarkably the importance and the options of a method that was idealized at the beginning as a quantization method quicker than the Dirac one.

The distribution of the issues obey the following pattern: in section 2 we carry out a brief review of the standard way to implement 
the Yang-Mills theory from the Maxwell theory.  In section 3 we begin the application of the FJ method obtaining the $SU(2)$ Yang-Mills starting from the Maxwell theory.  In section 4 we follow the same philosophy but now we obtain the $SU(2) \otimes U(1)$ Yang-Mills theory.  The conclusions are described in the last section.

\section{From Maxwell to Yang-Mills: the standard way}

The first published work to generalize the $U(1)$ Maxwell electromagnetic theory imposing non-Abelian group structure was introduced by Klein several decades ago \cite{Klein}. Years later, Yang and Mills were able to provide the full account of $SU(2)$ non-Abelian gauge theory \cite{Yang}. The basic idea, using electromagnetism as a guiding example, is to cancel the variation of the Dirac Lagrangian density
\be
\label{01}
{\cal L} = \bar{\psi}\(i \gamma^{\mu} \partial_{\mu} - m\) \psi 
\ee
\noindent under a local symmetry transformation of its fermion fields $\psi_{a}(x)$ in the intrinsic space
\be
\label{02}
\psi_{a}{\longrightarrow}\psi_{a}^{U}(x) = U_{ab}(x) \psi_{b}(x) ,
\ee
\noindent by the introduction of a gauge field $A_{\mu}^{a}(x)$. To achieve this, the derivative $\partial_{\mu} \equiv \frac{\partial}{\partial x^{\mu}}$ in the Lagrangian density (\ref{01}) is replaced by the covariant derivative
\be
\label{03}
D_{\mu} = \partial_{\mu} + g A_{\mu},
\ee
\noindent which transforms as $ D_{\mu}{\longrightarrow}UD_{\mu}U^{\dagger}$, whereas the gauge fields $A_{\mu}$ transform according to
\ba\label{04}
& &A_{\mu}(x)\equiv A_{\mu}^{a}(x) T^{a} \nonumber \\
& &{\longrightarrow}\;\;A_{\mu}^{U}(x) = U(x)A_{\mu}U^{\dagger}(x) + \frac{1}{g}U(x)\partial_{\mu}U^{\dagger}(x).
\ea

\noindent The subsequent interaction between gauge and matter fields is controlled by the coupling constant $g$.

The unitary group elements $U(x)\in SU(N)$,
\be
\label{05}
U(x) = \exp(-\alpha^{a}(x)T^{a}), \,\,\,\, \alpha \in R,
\ee
\noindent are generated by $(N^{2} - 1)$ anti-hermitian matrices $T_{a} = -T_{a}^{\dagger}$ that obey the Lie algebra
\be
\label{06}
[ T^{a}, T^{b}] = f^{abc} T^{c},
\ee
\noindent with the structure constants $f^{abc}$. The gauge fields $A_{\mu}$ live in a $(N^{2} - 1)$-dimensional space spanned by the generators $T^{a}$ with the additional operation of commutation (\ref{06}). By an infinitesimal expansion of (\ref{04}) in $\alpha$ using (\ref{06}), we have that $A_{\mu}$ must transform in the adjoint representation of $SU(N)$ as
\be
\label{07}
(A^{U})_{\mu}^{a}(x) = A_{\mu}^{a} + f^{abc} A_{\mu}^{b}\alpha^{c} + \frac{1}{g}\partial_{\mu}\alpha^{a} + {\it O}(\alpha^{2}),
\ee
\noindent where $(T^{b})^{ac}\equiv f^{abc}$. The covariant derivative in the adjoint representation is defined by
\be
\label{08}
D^{ab\mu} \equiv \delta^{ab} \partial^{\mu} + g f^{abc}A^{c\mu}.
\ee
\noindent And knowing the transformation properties of the gauge field (\ref{04}), one may construct from the gauge covariant field strength tensor
\be
\label{09}
F_{\mu \nu}^{a} = \partial_{\mu} A_{\nu}^{a} - \partial_{\nu} A_{\mu}^{a} + g f^{abc}A_{\mu}^{b}A_{\nu}^{c},
\ee
\noindent which can be derived by the commutator
\be
\label{010}
[D_{\mu}, D_{\nu}] = ig T^{a} F_{\mu \nu}^{a},
\ee
\noindent a gauge invariant contribution to the Lagrangian density. This contribution is given by 
\be
\label{011}
{\cal L} = - \frac{1}{4}F_{\mu \nu}^{a}F^{a \mu \nu}.
\ee
\noindent This is the Lagrangian density for the Yang-Mills theory, which is the starting point for all discussions concerning non-Abelian gauge theory.

\section{From $U(1)$ Maxwell Electromagnetic Theory to $SU(2)$ Yang-Mills Theory: the FJ way}

In order to present this methodology in details, let us consider again the $U(1)$ Maxwell theory in four dimensions, whose dynamics is governed by the following Lagrangian density
\be
\label{1}
{\cal L} = - \frac{1}{4} F_{\mu \nu}F^{\mu \nu},
\ee
\noindent where $F_{\mu \nu} = \partial_{\mu}A_{\nu} - \partial_{\nu}A_{\mu}$ and the space-time metric has the signature $g_{00} = + 1$ and $g_{ii} = -1$, with i = 1,2,3.

To accommodate an internal symmetry group, the original field is changed
\be
\label{2}
A_{\mu}{\longrightarrow}A_{\mu}^{a},
\ee
\noindent where ``{\it a}" denote an index relative to some internal symmetry group which we introduced by hand into the original theory. Thus, let us rewrite the original Lagrangian density as
\be
\label{3}
{\cal L} = - \frac{1}{4} G_{\mu \nu}^{a}G^{a \mu \nu},
\ee
\noindent where $G_{\mu \nu}^{a}$ is an arbitrary tensor which will be obtained later.
Notice that
\be
\label{4}
G_{\mu \nu}^{a}{\longrightarrow}F_{\mu \nu}^{a},
\ee 
\noindent if we reduce the new symmetry group to the original group.  Considering this condition, let us write $G_{\mu \nu}^{a}$ as,
\be
\label{5}
G_{\mu \nu}^{a} = F_{\mu \nu}^{a} + g {\tilde F}_{\mu \nu}^{a},
\ee
\noindent where $g$ is a parameter and ${\tilde F}_{\mu \nu}^{a} = {\tilde F}_{\mu \nu}^{a}(A_{\alpha}^{a})$ is an arbitrary antisymmetric tensor. Let us also suppose that the fields $A_{\mu}^{a}$ satisfy the non-Abelian algebra
\be
\label{6}
\[A_{\mu}^{a}, A_{\nu}^{b} \] = g \,\Sigma_{\mu \nu}^{a b} \,\Delta (x,y),
\ee
\noindent where $\Sigma_{\mu \nu}^{a b} = \Sigma_{\mu \nu}^{a b}(A_{\alpha}^{a})$ is also an arbitrary antisymmetric tensor and we will use conveniently from now on that $\Delta(x,y)=\delta^{(4)}(x-y)$. 

Now, in order to carry out the second step of the method, as we explained above, let us first rewrite the Lagrangian density in (\ref{3}) in first-order using the canonical momenta
\ba
\label{7}
\pi_{0}^{b} &=& 0, \\
\pi_{j}^{b} &=& \partial_{0}A_{j}^{b} + \partial_{j}A_{0}^{b} + g {\tilde F}_{0j}^{b}.
\ea
\noindent Thus, the first-order Lagrangian density has the following form
\ba
\label{8}
{\cal L} &=& \pi_{i}^{a} \dot{A}_{i}^{a} - \frac{1}{2}\pi_{i}^{a}\pi_{i}^{a} - \partial_{i}\pi_{i}^{a}A_{0}^{a} 
+ g {\tilde F}_{0i}^{a} \pi_{i}^{a} \nonumber\\
&-& \frac{1}{4} \(F_{ij}^{a}F^{aij} + 2g F^{aij}\tilde{F}^{aij} + g^2 \tilde{F}_{ij}^{a}\tilde{F}^{aij}\)\,\,. \nonumber \\
\mbox{}
\ea
Following the FJ symplectic method, the symplectic variables are
\be
\label{9}
\xi^{\alpha} = \(A_{i}^{a}, \pi_{i}^{a}, A_{0}^{a} \),
\ee
\noindent  and the two form matrix is given by
\be
\label{10}
f = \left(\begin{array}{ccc}
\frac{\delta \pi_{i}^{a}}{\delta A_{j}^{b}} - \frac{\delta \pi_{j}^{b}}{\delta A_{i}^{a}}  & -\delta_{ij}\delta^{ab}\Delta(x,y) & 0 \\
\delta_{ji}\delta^{ba}\Delta(x,y) & 0 & 0 \\
0& 0 & 0 
\end{array}\right).
\ee
It is obvious that this matrix is singular.  Also it has the following zero-mode $\nu = \(0 \ \ 0 \ \ 1\)$. This zero-mode leads us to the Gauss law constraint
\be
\label{11}
\Omega^a \equiv D^{abi} \pi_{i}^{b},
\ee
\noindent where $D^{abi}$ is the operator given by
\be
\label{12}
D^{abi} \equiv \delta^{ab} \partial^{i} + \frac{\delta {\tilde F}^{b0i}}{\delta A_{0}^{a}}. 
\ee
\noindent Following the prescription of the symplectic formalism, the constraint $\Omega^a$ is incorporated into the Lagrangian density to construct the new Lagrangian density as follows
\ba
\label{13}
{\cal L}' &=& \pi_{i}^{a} \dot{A}_{i}^{a} + \Omega^a \dot{\beta}^a - \Big[ \frac{1}{2}\pi_{i}^{a}\pi_{i}^{a} \nonumber \\
&+& \frac{1}{4} \(F_{ij}^{a}F^{aij} + 2g F^{aij}\tilde{F}^{aij} + g^2 \tilde{F}_{ij}^{a}\tilde{F}^{aij}\) \Big].
\ea

Now, the new symplectic variables are 
\be
\label{14}
\xi^{\alpha} = \(A_{i}^{a}, \pi_{i}^{a}, \beta^{a} \),
\ee
\noindent where $\beta$ is a Lagrange multiplier and the new two form matrix is
\begin{widetext}
\be
\label{15}
f = \left(\begin{array}{ccc}
\frac{\delta \pi_{j}^{b}}{\delta A_{i}^{a}} - \frac{\delta \pi_{i}^{a}}{\delta A_{j}^{b}}  & -\delta_{ij}\delta^{ab}\Delta(x,y) & \frac{\delta \Omega^{b}(y)}{\delta A_{i}^{a}(x)} \\
\delta_{ji}\delta^{ba}\Delta(x,y) & 0 & \(\delta^{ab} \partial_{i}^{y} + g \frac{\delta {\tilde F}_{0i}^{a}(y)}{\delta A_{0}^{b}(y)}\)\Delta(x,y) \\
- \frac{\delta \Omega^{a}(x)}{\delta A_{j}^{a}(y)}& - \(\delta^{ba} \partial_{j}^{x} + g \frac{\delta {\tilde F}_{0j}^{b}(x)}{\delta A_{0}^{a}(x)}\)\Delta(x,y) & 0 
\end{array}\right).
\ee
\end{widetext}

In order to have a gauge theory,  this matrix must be singular. To have, as the symmetry group of the theory, the $SU(2)$ Lie group, then we construct an ansatz for the zero-mode $\nu$, namely, 
\be
\label{16}
\nu = \(\delta^{da} \partial_{i}^{x} - g f^{adc}A_{i}^{c}(x),  \ \ - \frac{\delta \Omega^{d}(x)}{\delta A_{i}^{a}(y)}, \ \ - \delta^{da}\Delta(x,y)\).
\ee
\noindent In (\ref{16}), $f^{abc}$ are the structure constants of the $SU(2)$ group. The contraction of this 
zero-mode with the two form matrix in (\ref{15}) leads to the following relation
\be
\label{17}
\frac{\delta {\tilde F}_{0j}^{b}(x)}{\delta A_{0}^{a}(x)} = - f^{bdc}A_{j}^{c}(x).
\ee
\noindent After a straightforward calculation, we have that
\be
\label{18}
{\tilde F}_{0j}^{b}(x) = - f^{bdc}A_{0}^{d}(x)A_{j}^{c}(x).
\ee
\noindent Substituting this result in (\ref{11}), we obtain the following expression for the Gauss law constraint
\be
\label{19}
\Omega^a(y) = \partial^{j} \pi_{j}^{a}(y) - g f^{cba}A_{j}^{b}(y) \pi_{j}^{c}(y).
\ee

Since the theory must remain covariant, the introduction of (\ref{18}) in (\ref{5}) permit us to conclude that the tensor $G_{\mu \nu}^{a}$ has the general form
\be
\label{20}
G_{\mu \nu}^{a} = F_{\mu \nu}^{a} - g f^{abc}A_{\mu}^{b} A_{\nu}^{c}.
\ee

If we reduce the $SU(2)$ group to $U(1)$ group, making $g{\longrightarrow}0$ in (\ref{16}), we obtain the field transformations of the $U(1)$ Maxwell theory. Therefore, we can write (\ref{20}) as  
\be
\label{21}
G_{\mu \nu}^{a} = \partial_{\mu}A_{\nu}^a - \partial_{\nu}A_{\mu}^a - g f^{abc}A_{\mu}^{b} A_{\nu}^{c}.
\ee

In our final discussion, substituting (\ref{18}) in (\ref{12}) we obtain the expression for the operator $D^{ab\mu}$, i. e.,
\be
\label{22}
D^{ab\mu} \equiv \delta^{ab} \partial^{\mu} + g f^{abc}A^{c\mu}, 
\ee
\noindent which permit us to define the covariant derivative
\be
\label{23}
D_{\mu} \equiv \partial_{\mu} + i g A_{\mu}^{a}T^{a}, 
\ee
\noindent where $T^{a}$ are the generators of the algebra in a standard representation of the group. As well known, the advantage of introducing the covariant derivative is that it transforms covariantly under gauge transformations generated
by zero-mode. This means that the action will be gauge invariant if we replace $\partial_\mu$ by $D_\mu$. Note that here the covariant derivative arises naturally from the methodology of the symplectic formalism.

\section{From $U(1)$ Maxwell Electromagnetic Theory to $SU(2)\otimes U(1)$ Yang-Mills Theory: the FJ way}

To construct a gauge theory based on the $SU(2)\otimes U(1)$ group, let us add to the Lagrangian density (\ref{01}) the following term,
\be
\label{24}
{\cal L} = - \frac{1}{4} G_{\mu \nu}^{a}G^{a \mu \nu}, 
\ee
\noindent where $G_{\mu \nu}^{a}$ is an arbitrary tensor which will be obtained later and ``{\it a}" denote an index relative to the symmetry non-Abelian group $SU(2)$. As in the previous section, we begin with the $U(1)$ Maxwell electromagnetic theory.
In this way, let us construct a $G_{\mu \nu}^{a}$ tensor with the form
\be
\label{25}
G_{\mu \nu}^{a} = \partial_{\mu}A_{\nu}^a - \partial_{\nu}A_{\mu}^a + g_{w} {\tilde W}_{\mu \nu}^{a}\,\,, 
\ee
\noindent where $g_{w}$ is a parameter and ${\tilde W}_{\mu \nu}^{a} = {\tilde W}_{\mu \nu}^{a}(W_{\alpha}^{b})$ is an arbitrary antisymmetric tensor. Let us also suppose that the fields $W_{\mu}^{a}$ satisfy the non-Abelian algebra
\ba\label{26}
\[W_{\mu}^{a}, W_{\nu}^{b} \] = g_{w} \Sigma_{\mu \nu}^{a b} \Delta(x,y)\,\,, \\
\mbox{} \nonumber
\ea
\noindent where $\Sigma_{\mu \nu}^{a b} = \Sigma_{\mu \nu}^{a b}(W_{\alpha}^{c})$ is also an arbitrary antisymmetric tensor.
Thus, we have the following Lagrangian density
\be
\label{27}
{\cal L} = - \frac{1}{4} F_{\mu \nu}F^{\mu \nu} - \frac{1}{4} G_{\mu \nu}^{a}G^{a \mu \nu}, 
\ee
which in the first-order form it is given by
\ba
\label{28}
{\cal L} &=& \pi_{i}\dot{B}_{i} + \pi_{i}^{a} \dot{W}_{i}^{a} - \frac{1}{2}\pi_{i}\pi_{i} - \partial_{i}\pi_{i}B_{0}\nonumber \\ 
&-&\frac{1}{4}F_{ij}F^{ij} - \frac{1}{2}\pi_{i}^{a}\pi_{i}^{a} - \partial_{i}\pi_{i}^{a}W_{0}^{a} 
+ g_{w} {\tilde W}_{0i}^{a} \pi_{i}^{a} \\
&-&\frac{1}{4} \(W_{ij}^{a}W^{aij} + 2g_{w} W^{aij}\tilde{W}^{aij} + g_{w}^2 \tilde{W}_{ij}^{a}\tilde{W}^{aij}\)\,\,,, \nonumber 
\ea
\noindent where $B_\mu$ is the Maxwell field and
\ba
\label{29}
\pi_{0} &=& 0, \nonumber \\
\pi_{j} &=& \partial_{0}B_{j} + \partial_{j}B_{0}, \nonumber \\
\pi_{0}^{b} &=& 0, \nonumber \\
\pi_{j}^{b} &=& \partial_{0}W_{j}^{b} + \partial_{j}W_{0}^{b} + g_{w} {\tilde W}_{0j}^{b},
\ea
\noindent are the canonical momenta. The symplectic variables are
\be
\label{30}
\xi^{\alpha} = \(B_{i}, \pi_{i}, B_{0}, W_{i}^{a}, \pi_{i}^{a}, W_{0}^{a} \),
\ee

\noindent and the two form matrix is given by

\begin{widetext}
\be
\label{31}
f = \left(\begin{array}{cccccc}
0 & 0 & -\delta_{ij}\Delta(x,y) & 0 & 0 & 0 \\
0 & \delta_{ji}\Delta(x,y) & 0 & 0 & 0 & 0 \\
0 & 0 & 0 & 0 & 0 & 0 \\
0 & 0 & 0 & \frac{\delta \pi_{i}^{a}}{\delta W_{j}^{b}} - \frac{\delta \pi_{j}^{b}}{\delta W_{i}^{a}} & -\delta_{ij}\delta^{ab}\Delta(x,y) & 0 \\
0 & 0 & 0 & \delta_{ji}\delta^{ba}\Delta(x,y) & 0 & 0 \\
0 & 0 & 0 & 0 & 0 & 0 
\end{array}\right).
\ee
\end{widetext}

\noindent It is obvious that this matrix is singular and it has the following zero-modes
\ba
\label{32}
\nu &=& \(0 \ \ 0 \ \ 1 \ \ 0 \ \ 0 \ \ 0 \), \nonumber \\
\eta &=& \(0 \ \ 0 \ \ 0 \ \ 0 \ \ 0 \ \ 1 \),
\ea
\noindent which lead to the constraints
\ba
\label{33}
\omega \equiv \partial^{i} \pi_{i}, \nonumber \\
\Omega^a \equiv D^{abi} \pi_{i}^{b},
\ea
\noindent where $D^{abi}$ is the operator
\be
\label{34}
D^{abi} \equiv \delta^{ab} \partial^{i} + g_{w} \frac{\delta {\tilde F}^{b0i}}{\delta A_{0}^{a}}. 
\ee

Following the prescription of the symplectic formalism, the constraints are incorporated into the Lagrangian 
density to construct a new Lagrangian density as follows

\ba
\label{35}
{\cal L} &=& \pi_{i} \dot{B}_{i} + \pi_{i}^{a} \dot{W}_{i}^{a} + \omega \dot{\lambda} + \Omega^a \dot{\beta}^a - 
\Big[ \frac{1}{2}\pi_{i} \pi_{i} \,+\, \frac{1}{4} F_{ij}F^{ij}  \nonumber \\
&+&  \frac{1}{4} \(W_{ij}^{a}W^{aij} + 2g_{w} W^{aij}\tilde{W}^{aij} + g_{w}^2 \tilde{W}_{ij}^{a}\tilde{W}^{aij}\) \Big] \,\,, \nonumber  \\
\mbox{}
\ea
\noindent where $\lambda$ and $\beta^{a}$ are Lagrange multipliers.
The new symplectic variables are
\be
\label{36}
\xi^{\alpha} = \(B_{i}, \pi_{i}, \lambda, W_{i}^{a}, \pi_{i}^{a}, \beta^{a} \),
\ee
\noindent and the new two form matrix is
\begin{widetext}
\ba
\label{37}
f = \left(\begin{array}{cccccc}
0 & -\delta_{ij}\Delta(x,y) & 0 & 0 & 0 & 0 \\
\delta_{ji}\Delta(x,y) & 0 & - \partial_{i}^{y}\Delta(x,y) & 0 & 0 & 0 \\
0 & \partial_{j}^{x}\Delta(x,y) & 0 & 0 & 0 & 0 \\
0 & 0 & 0 & \frac{\delta \pi_{j}^{b}}{\delta W_{i}^{a}} - \frac{\delta \pi_{i}^{a}}{\delta W_{j}^{b}} & -\delta_{ij}
\delta^{ab}\Delta(x,y) & \frac{\delta \Omega^{b}(y)}{\delta W_{i}^{a}(x)} \\
0 & 0 & 0 & \delta_{ji}\delta^{ba}\Delta(x,y) & 0 & \(\delta^{ab} \partial_{i}^{y} + g_{w} \frac{\delta {\tilde W}_{0i}^{a}(y)}{\delta W_{0}^{b}(y)}\)\Delta(x,y) \\
0 & 0 & 0 & - \frac{\delta \Omega^{a}(x)}{\delta W_{j}^{b}(y)} & - \(\delta^{ba} \partial_{j}^{x} + g_{w} \frac{\delta {\tilde W}_{0j}^{b}(y)}{\delta W_{0}^{a}(x)}\)\Delta(x,y) & 0 
\end{array}\right)\,\,.\nonumber \\
\mbox{}
\ea
\end{widetext}


Since this new theory must be a gauge theory based on the $SU(2)\otimes U(1)$ group, then this matrix must be singular and we construct an ansatz for the zero-modes, namely,
\ba
\label{38}
& &\nu \,=\, \(- \partial_{i} \ \ 0 \ \ 1 \ \ 0 \ \ 0 \ \ 0 \), \nonumber \\
& &\eta \,=\, \\
& & \(0 \ \ 0 \ \ 0 \ \ \delta^{ab}\partial_{i}^{x} - g_{w} f^{abc} W_{i}^{c}(x), \ - \frac{\delta \Omega^{a}(x)}{\delta W_{i}^{b}(y)}, \ \delta^{ab}\Delta(x,y)\) \nonumber 
\ea
\noindent The zero-mode $\nu$ does not generate any new constraint and the same will occur with $\eta$ since
\be
\label{39}
\frac{\delta {\tilde W}_{0j}^{b}(x)}{\delta W_{0}^{d}(x)} = - f^{bdc}W_{j}^{c}(x),
\ee
\noindent and after a straightforward calculation, we have that
\be
\label{40}
{\tilde W}_{0j}^{b}(x) = - f^{bdc}W_{0}^{d}(x)W_{j}^{c}(x).
\ee

Since the theory must remain covariant, we substitute equation (\ref{40}) in (\ref{25}) so that now the tensor $G_{\mu \nu}^{a}$ has the general form
\be
\label{41}
G_{\mu \nu}^{a} = \partial_{\mu}W_{\nu}^a - \partial_{\nu}W_{\mu}^a - g_{w} f^{abc}W_{\mu}^{a} W_{\nu}^{c}.
\ee

In agreement with the FJ symplectic formalism, the zero-modes $\nu$ and $\eta$ are the generators of the infinitesimal gauge transformation of the action (39), given by
\ba
\label{42}
\delta B_{\mu} &=& - \partial_{\mu} \epsilon,\nonumber \\
\delta W_{\mu}^{a} &=& \partial_{\mu} \epsilon^{a} - g_{w} f^{abc} W_{\mu}^{b}\epsilon^{c}\,\,,
\ea

\noindent and we can realize another advantage of the FJ formalism concerning the gauge invariance of the system.  In these last lines, we could see that the zero mode, which is an item of the process, helps in the construction of the gauge symmetries of the theory.  With these results we are led to think that the arbitrariness of the zero mode can lead us to construct a whole family of gauge symmetries and gauge theories.  However, as shown in recent papers \cite{Oliveira}, the choice of the zero-mode is connected with the physical features of the system.  In other words, a family of gauge symmetries can still be constructed but the physical coherence must be carefully  analyzed for each zero mode introduced in the formalism.

\section{conclusion}

In this work we have presented an completely new application of the FJ method which main purpose, in this work, is to obtain non-Abelian field theories. This approach is based on the so-called FJ symplectic formalism, which does not classify the constraints of the theory as first or second class neither as primary or secondary constraints.   To carry out this  goal, we follow two steps. 

In the first step, the original Abelian fields are modified in order to introduce the non-Abelian algebra.  In the second step, the FJ symplectic formalism is implemented and the gauge symmetry relative to some  non-Abelian symmetry group is introduced. This was accomplished through the zero-mode of the two form matrix. In other words we can say that, obeying the FJ symplectic formalism, the two form  matrix must be singular.  This last condition helps us to construct a convenient zero-mode that contracts with the matrix resulting in a Lagrangian that has the non-Abelian symmetry group described just below.

Here, our starting point was the $U(1)$ Maxwell electromagnetic theory. To obtain the $SU(2)$ and $SU(2)\otimes U(1)$ Yang-Mills theories we choose convenient zero-modes for the two form matrix.  As a very convenient feature of the FJ formalism, the zero-mode is the generator of the infinitesimal gauge transformation, but it is, at the same time, connected to the physical characteristics of the system.  In our case, the objective was to obtain a non-Abelian theory and in this sense we can interpret the FJ formalism as a kind of mapping between the original action and the final one.  However, we must not forget that the method was conceived with the purpose of quantization in a quicker way in comparison with the Dirac method.  On the other side we can always understand this mapping as a sort of duality.

\section{ Acknowledgments}
This work is supported in part by FAPEMIG and CNPq, Brazilian
Research Agencies. EMCA, CN, WO and CW would like to acknowledge
the CNPq and ACRM would like to acknowledge the FAPEMIG.  EMCA would like to thank the kindness and hospitality of Departamento de F\' isica of Universidade Federal de Juiz de Fora, where part of this work has been carried out.

\end{document}